\documentclass[11pt,twoside]{article}
\usepackage{asp2010}

\input epsf
\usepackage{graphicx}

\resetcounters

\begin{document}

\title{Future prospects for inference on solar-type stars}
\author{W. J. Chaplin \affil{School of Physics and Astronomy,
University of Birmingham, Edgbaston, Birmingham, B15 2TT, UK}}

\begin{abstract}

We discuss prospects for asteroseismic inference on solar-type stars,
in particular opportunities that are being made possible by the large
ensemble of exquisite-quality \emph{Kepler} data.

\end{abstract}

 \section{Introduction}
 \label{sec:intro}

The observational basis for stellar physics is being enhanced
significantly by a new era of satellite and ground-based telescope
observations of unprecedented quality and scope.  The recent launch of
the NASA \emph{Kepler Mission} has meant a huge breakthrough in amount
and quality of data for the study of solar-type stars using the
techniques of asteroseismology (Gilliland et al. 2010, Chaplin et
al. 2010).

\begin{figure*}
 \centerline
 {\epsfxsize=9.0cm\epsfbox{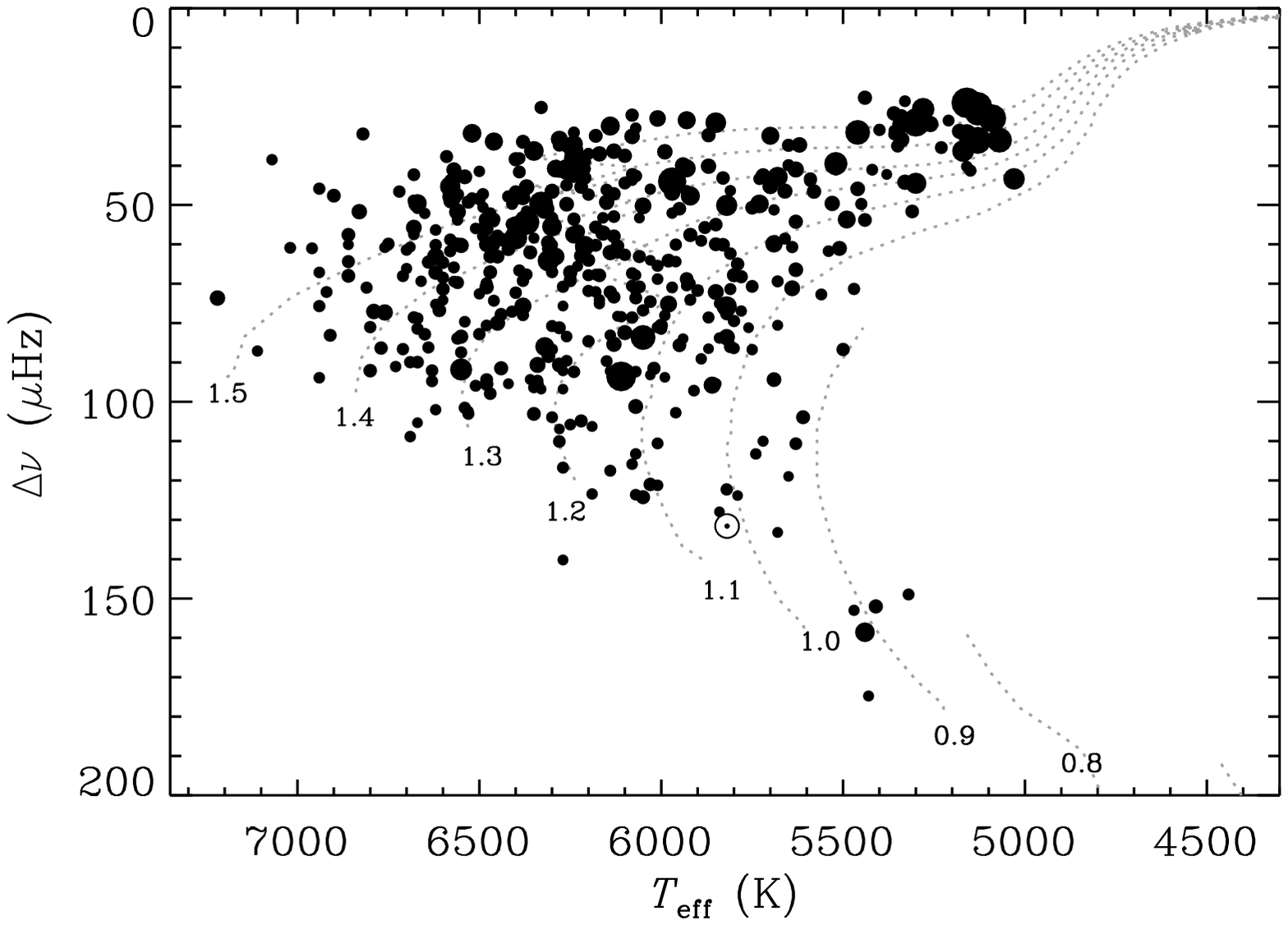}}
 \centerline
 {\epsfxsize=9.0cm\epsfbox{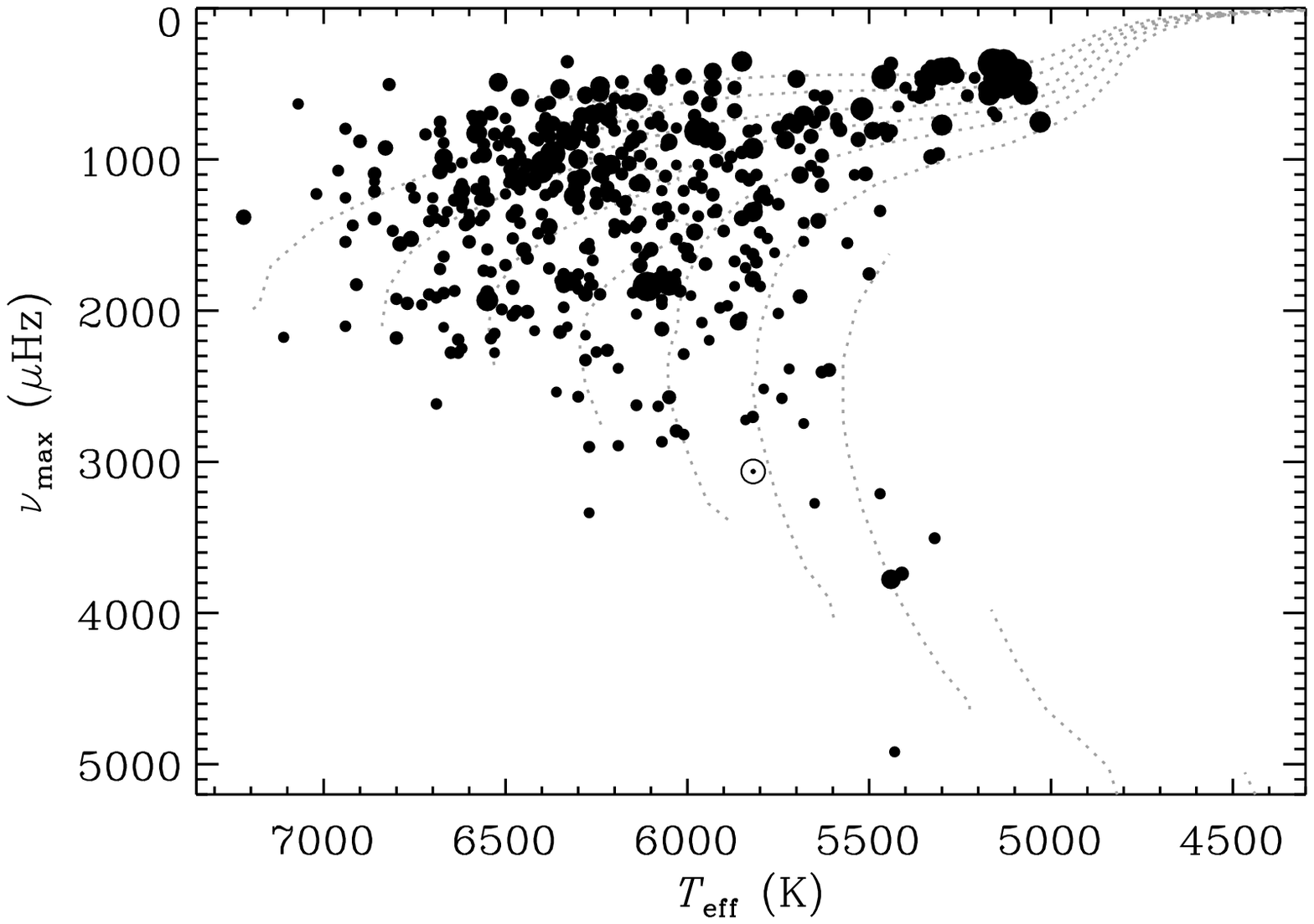}}
 \centerline
 {\epsfxsize=9.0cm\epsfbox{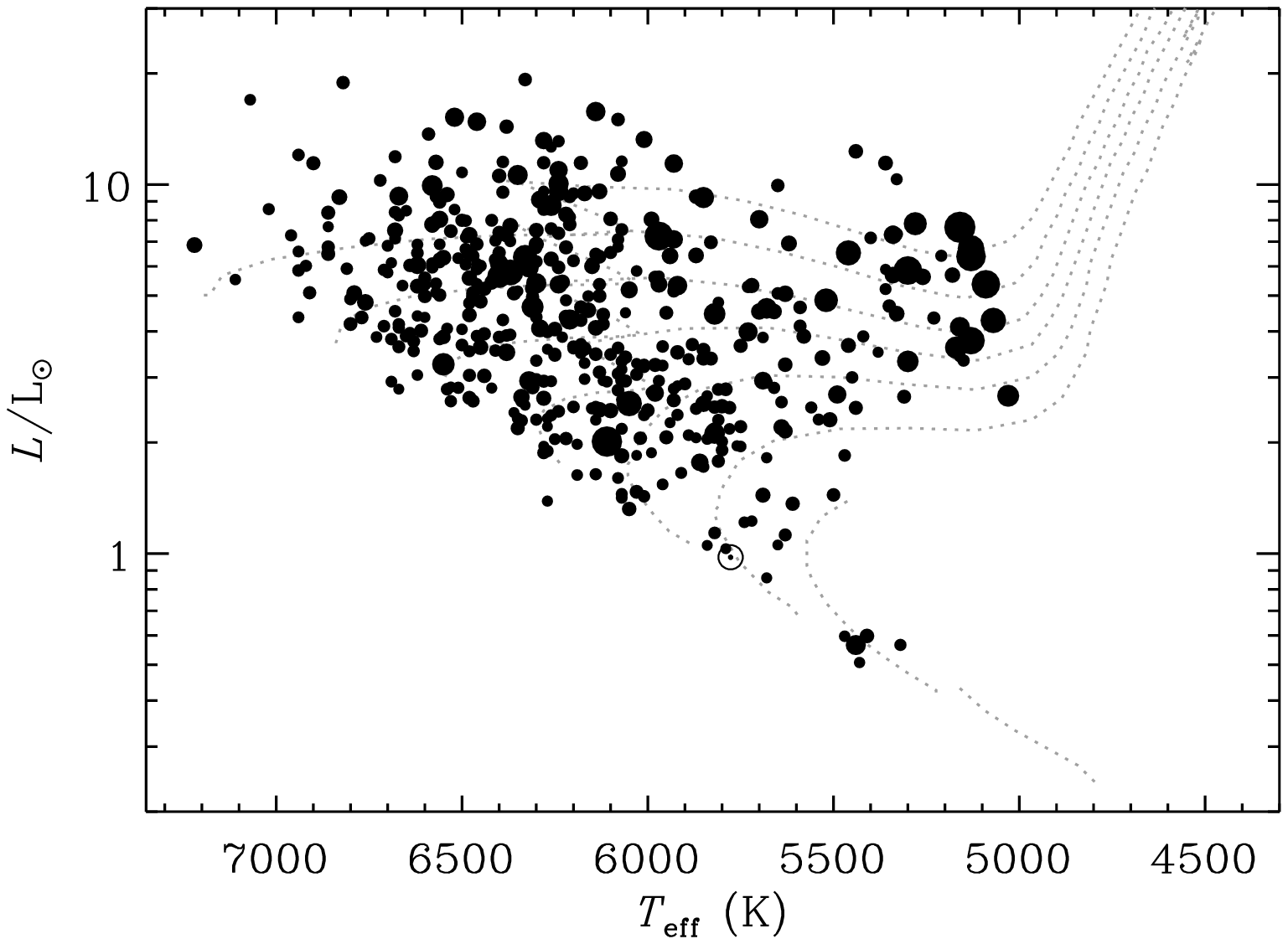}}

  \caption{Results from \emph{Kepler} on solar-type stars with
  detectable oscillations. Plotted, versus effective temperature
  $T_{\rm eff}$, are: Observed average large frequency separations,
  $\Delta\nu$ (top panel); observed frequencies of maximum oscillations
  power, $\nu_{\rm max}$ (middle panel); observed luminosities $L
  \propto R^2T_{\rm eff}^4$, computed from the seismically determined
  radii, $R$ (bottom panel). Symbol sizes are proportional to observed
  $\rm SNR_{\rm tot}$. The location of the Sun is marked with the
  usual solar symbol. The dotted lines are evolutionary tracks (Padova
  models) for solar composition, computed for masses ranging from 0.7
  to $1.5\,\rm M_{\odot}$ (Girardi et al. 2002, 2004; Marigo et
  al. 2008).}

 \label{fig:ensemble}
\end{figure*}

Thanks to \emph{Kepler} the number of solar-type stars with good
seismic observations has increased from about 20 to several-hundred
(Chaplin et al. 2011a; see Fig.~\ref{fig:ensemble}). The French-led
CoRoT satellite (Michel et al. 2008, Appourchaux et al. 2008) also
continues to provide high-quality asteroseismic data on a smaller
number of solar-type stars. It is based on these dramatic improvements
in data quality and quantity that exciting new results from
asteroseismology are feeding into studies in stellar physics,
exoplanets, galactic and extra-galactic physics and cosmology.

 \section{Overview: opportunities for seismic analysis of solar-type
  stars}
 \label{sec:overview}

Solar-type stars have sub-surface convection zones and, like the Sun,
display solar-like oscillations.  The rich information content of
these seismic signatures means that the fundamental stellar properties
(e.g., mass, radius, and age) may be measured and the internal
structures constrained to levels that would not otherwise be possible
(e.g., see Gough 1987; Cunha et al. 2007; Aerts et al. 2010). While
helioseismology has shown that we can model $\sim 1\,\rm M_{\odot}$
stars quite well, we do not yet now if we can model more massive stars
with convective cores properly (the sizes of the convective cores, and
the amount of core overshoot, affect the main-sequence lifetimes of
the stars; e.g., see Cunha \& Metcalfe 2007). Multi-month observations
of solar-type targets with \emph{Kepler} will allow us to obtain
accurate and precise estimates of the basic parameters of
\emph{individual modes}, such as frequencies, frequency splittings,
amplitudes, damping rates, and asymmetries of the resonant peaks,
covering many radial orders.

Multi-month and multi-year data on a sizable ensemble of solar-like
oscillators opens the possibility to:
 \begin{itemize}

 \item study the internal rotation rates of solar-type stars, compare
 internal and surface rates of rotation, test theories of angular
 momentum evolution, and test and calibrate gyrochronology;

 \item measure the helium abundances, and signatures of convective
 cores, to test theories of stellar evolution;

 \item measure the depths of the sub-surface convective envelopes,
 signatures of stellar cycles and the acoustic asphericities, to test
 dynamo theories; and

 \item study the interaction of convection with the solar-like
 oscillations, using information from the frequency dependence of mode
 amplitudes, mode damping rates and mode peak asymmetries;

 \end{itemize}


Use of individual frequencies increases the information content
provided by the seismic data (e.g., see Monteiro et al. 2000). The
frequencies offer the potential for inferring the radial structure
inside stars (e.g., Basu et al. 2001; Roxburgh \& Vorontsov 2003);
they also allow detection of signatures of regions of abrupt
structural change in the stellar interiors, e.g., the near-surface
ionization zones and the bases of the convective envelopes (see Houdek
\& Gough 2007, and references therein). Inference on the ionization
zone signatures in principle allows constraints to be placed on the
envelope He abundances (e.g., Mazumdar et al. 2006); while estimation
of the depths of the convective envelopes will provide key information
for the dynamo modellers.

Asteroseismology can provide precise estimates of ages of solar-type
stars. Inference on ages of course relies on stellar evolutionary
theory.  The stellar models must include all of the requisite physics
that we consider to be significant in determining the evolutionary
status of the star, and as a result its observable
properties. Inaccuracies in the descriptions of these quantities may
of course lead to systematic errors in estimates of the fundamental
stellar properties. There are uncertainties related to modelling of,
for example, convective motions and convective overshooting at the
convective boundaries, microscopic diffusion, and the impact of
dynamical processes such as instabilities and flows associated with
internal rotation. It will be possible to use precise data on the
oscillation frequencies to constrain the input physics to the stellar
evolutionary models.

With asteroseismic ages in hand it will, for example, be possible to
validate methods of ageing stars based on the observed surface rates
of rotation (gyrochronology; see Barnes 2003). These surface
signatures may be detected, and then quantified, in the \emph{Kepler}
and CoRoT lightcurves. From the long datasets it is possible to
extract estimates of frequency splittings of non-radial modes, which
have contributions from internal rotation and magnetic fields. This
opens the possibility to study the relationship between the mean
internal rotation and the surface rates of rotation in the solar-type
ensemble. Measurement of asymmetries of the frequency splittings also
allows constraints to be placed on the surface distributions of active
regions (e.g., see Chaplin et al. 2007; Chaplin 2011), results that
may be compared with inferences drawn from modelling of the surface
signatures of rotation.  And it is also possible to estimate stellar
angles of inclination from measurement of the amplitude ratios of
components of non-radial modes (e.g., see Gizon \& Solanki 2003;
Ballot et al. 2006).

The observed oscillation modes in these solar-type stars are
intrinsically stable (e.g. Balmforth 1992a; Houdek et al. 1999) but
driven stochastically by the vigorous turbulence in the superficial
stellar layers (e.g. Goldreich \& Keeley 1977; Balmforth 1992b; Samadi
\& Goupil 2001).  Solar-like mode peaks have an underlying form that
follows, to a reasonable approximation, a Lorentzian. The widths of
the Lorentzians provide a measure of the linear damping rates, while
the amplitudes are determined by the delicate balance between the
excitation and damping. Small departures from the Lorentzian form --
which makes the peaks asymmetric -- provide information on the
location and properties of the acoustic sources, and correlation of
the oscillations with the stellar granulation. Accurate and precise
measurement of these various parameters, and their variation in
frequency, provides the means to infer various important properties of
the still poorly understood near-surface convection.

Further inferences on near-surface physics may potentially be acquired
by differential seismic analyses of simultaneous observations of the
same star made in photometry and Doppler velocity. For example, there
is the exciting prospect of our having simultaneous data on the
solar-type binary 16~Cyg from observations made in photometry by
\emph{Kepler} and in Doppler velocity by the Stellar Observations
Network Group (SONG) (Grundahl et al. 2009). The opportunities for
collecting simultaneous data in photometry and Doppler velocity may be
limited in the near future to only a small number of stars, and as
such it also behoves us to look for twins of \emph{Kepler} and CoRoT
stars in data already collected by ground-based telescopes.

Finally, extended observations of solar-type targets spanning several
years will allow us to ``sound'' stellar activity cycles, by detecting
systematic changes in mode parameters such as frequencies, amplitudes
and damping rates (e.g., see Chaplin et al. 2007; Metcalfe et
al. 2007; Karoff et al. 2009).

 \section{Ensemble studies}
 \label{sec:ensemble}

\emph{Kepler} is opening exciting new opportunities to conduct
\emph{ensemble seismology} thanks to the large number of stars being
observed. The large, homogeneous \emph{Kepler} ensemble will for the
first time allow a seismic survey of a population of solar-type field
stars to be made. A statistical survey of trends in important seismic
parameters will allow tests of basic scaling relations, comparisons
with trends predicted from modelling, and lead to important insights
on the detailed modelling of stars. This work is now in progress (e.g.,
see Chaplin et al. 2011b, Huber et al. 2011, Verner et al. 2011, White
et al. 2011a)

One may also pick from a large ensemble pairs, small groups or
sequences of stars that share common stellar properties, e.g., mass,
composition, or surface gravity, e.g., Silva-Aguirre et al. (2011a)
demonstrate selection of a $1\,\rm M_{\odot}$ sequence of
\emph{Kepler} stars. With selected data of this type we can perform
what one might term \emph{differential} (or \emph{comparative})
seismology of stars, e.g., in analysing the selected stars one may
eliminate or suppress any dependence of the modelling or results on
the common property, or properties. By selecting, for example, a
sequence of stars of very similar mass and composition it is possible
to produce an exceedingly accurate and robust relative age
calibration, and give the potential to map evolutionary sequences of
internal properties and structures, allowing exquisite tests of
stellar evolutionary models.  By selecting stars with very similar
surface gravities, one may potentially probe differences in
near-surface physics and convection.

 \section{Inference on regions of abrupt structural change in stellar
   interiors}
 \label{sec:abrupt}

Regions of stellar interiors where the structure changes abruptly,
such as the boundaries of convective regions, give rise to departures
from the regular frequency separations implied by an asymptotic
description. Allowance must be made for the signatures of these
changes to in principle avoid biases in the inferred fundamental
properties. This might seem like an unwanted complication. However,
careful measurement of these signatures not only brings cleaner
inference on the properties -- i.e., the signatures can first be
removed from the mode frequencies before the direct fitting is
performed -- but it also elucidates other important parameters and
physical properties of the stars. There are signatures left by the
ionization of helium in the near-surface layers of the stars.
Measurements of these signatures allow tight constraints to be placed
on the helium abundance, something that would not otherwise be
possible in such cool stars (because the ionization temperatures are
too high to yield usable photospheric lines for spectroscopy). And as
noted above, there are also signatures left by the locations of
convective boundaries. It is therefore possible to pinpoint the lower
boundaries of convective envelopes. These regions are believed to play
a key r\^ole in stellar dynamos. Furthermore, it is also possible to
estimate the sizes of convective cores. Measurement of the sizes of
these cores, and the overshoot of the convective motions into the
layers above, is important because it can provide an even more
accurate calibration of the ages of the affected stars.  The mixing
implied by the convective cores, and the possibility of mixing of
fresh hydrogen fuel into the nuclear burning cores -- courtesy of the
regions of overshoot -- affects the main-sequence lifetimes (e.g., see
Mazumdar et al. 2006).

The characteristics of the signatures imposed on the mode frequencies
depend on the properties and locations of the regions of abrupt
structural change. When the regions lie well within the acoustic
cavities, a periodic component is manifest in the frequencies. The
period of the signature relates to the acoustic radius of the region,
while the amplitude of the signature provides a measure of the size of
the effect. Two signatures of this type have already been well studied
in the solar case: one due to the discontinuity in the gradient of the
sound speed at the base of the convective envelope; and another due to
changes in the adiabatic exponent in the near-surface helium
ionization zones.

While the periodic signatures from the helium ionization zones may
already be readily apparent in the large frequency separations, their
signals may be better isolated by, for example, taking second
differences of frequencies of modes having the same angular degree $l$
(e.g., Basu et al. 2004), or by subtracting the frequencies from a
smoothly varying function in the overtone number, $n$ (Verner, Chaplin
\& Elsworth 2006). The signature from the base of the convective
envelope is also apparent in the second differences, although at a
reduced amplitude compared to the helium signature. Better diagnostics
from which to extract the convective-envelope signature are instead
frequency differences that make use of the $l=0$ and $l=1$ modes
(Roxburgh 2009). In the p-mode frequency spectrum, modes of odd
angular degree lie about halfway between modes of even angular
degree. The diagnostics are provided by the deviations of the $l=0$
and $l=1$ mode frequencies from the exact halfway frequencies.

The key to extracting these signatures in other solar-type stars is to
have sufficient precision in estimates of the frequencies. Basu et
al. (2004) have shown that it should be possible to extract the helium
signatures from solar-type main-sequence stars provided the
frequencies are estimated to a precision of at least $\approx 1$ part
in 10,000.  Similar constraints are imposed in respect of the
measurement of the signatures of the bases of the convective envelopes
(e.g., see the tests undertaken by Ballot, Turck-Chi\'eze \& Garc\'ia
2004, and Verner, Elsworth \& Chaplin 2006). Extraction of the helium
signatures allows an estimate to be made of the helium abundance in
the stellar envelopes. This should be possible to a precision of 1 to
2\,per cent. While extraction of the convective-envelope signature
provides a well-constrained estimate of the acoustic radius of the
envelope, and also allows constraints to be placed on the overshoot
into the radiatively stratified layer below (e.g., see Monteiro,
Christensen-Dalsgaard \& Thompson 2000).

When regions of abrupt structural change do not lie well within the
mode cavities, the signatures they leave are more subtle. This is the
case for the signatures left by small convective cores found in
solar-type stars that are slightly more massive than the Sun (Cunha \&
Metcalfe 2007).  Suitable combinations of the mode frequencies can
provide sensitive diagnostics not only of the presence of convective
cores, but also of their size (Cunha \& Brand\~ao 2011; Silva-Aguirre
et al. 2011b). Soriano \& Vauclair (2008) have also shown how the
abrupt gradients at the edges of the convective cores can cause the
small frequency separations to reverse their sign. Measurement of the
separations can therefore also serve to elucidate the presence, and
then constrain the properties of, the cores.

 \section{Diagnostic potential of modes of mixed character}
 \label{sec:mixed}

Next, we touch on the diagnostic potential of avoided crossings (Osaki
1975; Aizenman et al. 1977) in solar-type stars.  These avoided
crossings are a tell-tale indicator that the stars have evolved
significantly. In young solar-type stars there is a clear distinction
between the frequency ranges that will support acoustic (pressure, or
p) modes and buoyancy (gravity, or g) modes.  As stars evolve, the
maximum buoyancy (Brunt-V\"ais\"al\"a) frequency increases. After
exhaustion of the central hydrogen, the buoyancy frequency in the deep
stellar interior may increase to such an extent that it extends into
the frequency range of the high-order acoustic modes. Interactions
between acoustic modes and buoyancy modes may then lead to a series of
avoided crossings, which result from the frequencies being ``bumped'',
with the affected modes taking on mixed p and g
characteristics. Measurement of the frequency signatures of these
avoided crossings has the potential to provide exquisite constraints
on the fundamental stellar properties (Bedding 2011).  Very little
data had been available historically. Observational evidence for
avoided crossings had been uncovered in ground-based asteroseismic
data on two bright stars, $\eta$\,Boo and $\beta$~Hyi
(Christensen-Dalsgaard et al. 1995; Kjeldsen et al. 1995; Bedding et
al. 2007; Do\u{g}an et al. 2010).  Deheuvels et al. (2010) also
recently reported evidence for avoided crossings in CoRoT observations
of the G-type star HD49385, with Deheuvels \& Michel (2010) using an
elegant analysis based on coupled oscillators to discuss the results.
\emph{Kepler} now promises dramatic changes in this area, because
within the large \emph{Kepler} ensemble of solar-like oscillators
there is a clear selection of stars showing avoided crossings.  One of
the three solar-like stars selected for the first \emph{Kepler} paper
on solar-like oscillators -- KIC~11026764 -- is a beautiful case in
point, and has been the subject of further in-depth study (Metcalfe et
al. 2010a). The data provided by \emph{Kepler} will by necessity drive
developments in analysis to allow the diagnostic potential of avoided
crossings to be fully exploited, work that is already in progress
(e.g., see discussion of the new so-called p-g diagram in Bedding
2011, and White et al. 2011b)

 \section{Seismic diagnostics of stellar cycles}
 \label{sec:cycle}

The availability of long timeseries data on solar-type stars, courtesy
of the \emph{Kepler} and CoRoT is now making it possible to ``sound''
stellar cycles with asteroseismology. The prospects for such studies
have been considered in some depth (e.g., Chaplin et al. 2007, 2008;
Metcalfe et al. 2008; Karoff et al. 2009), and in the last year the
first convincing results on stellar-cycle variations of the p-mode
frequencies of a solar-type star (the $F$-type star HD49933) were
reported by Garc\'ia et al. (2010). This result is important for two
reasons: first, the obvious one of being the first such result,
thereby demonstrating the feasibility of such studies; and second, the
period of the stellar cycle was evidently significantly shorter than
the 11-yr period of the Sun (probably between 1 and 2\,yr). If other
similar stars show similar short-length cycles, there is the prospect
of being able to ``sound'' perhaps two or more complete cycles of such
stars with \emph{Kepler} (assuming the mission is extended, as
expected, to 6.5\,yr or more). The results on HD49933 may be
consistent with the paradigm that stars divide into two groups,
activity-wise, with stars in each group displaying a similar number of
rotation periods per cycle period (e.g., see B\"ohm-Vitense 2007),
meaning solar-type stars with short rotation periods -- HD49933 has a
surface rotation period of about 3\,days -- tend to have short cycle
periods. We note that Metcalfe et al. (2010b) recently found another
$F$-type star with a short (1.6\,yr) cycle period (using chromospheric
H \& K data). Extension of the \emph{Kepler} Mission will of course
also open the possibility of detecting full swings in activity in
stars with cycles having periods up to approximately the length of the
solar cycle.

\acknowledgements The author would like to thank H.~Shibahashi for the
opportunity to contribute this review. He also acknowledges the UK
STFC for grant funding to support his research in asteroseismology.


\begin{thebibliography}{}


\bibitem[]{} Aerts, C., Christensen-Dalsgaard, J., Kurtz, D. W., 2010,
{\it Asteroseismology}, Springer, Heidelberg

\bibitem[]{} Aizenman, M., Smeyers, P., Weigert, A., 1977, A\&A, 58, 41

\bibitem[]{} Appourchaux, T., Michel, E., Auvergne, M., et al., 2008,
A\&A, 488, 705

\bibitem[]{} Ballot, J., Garc\'ia, R. A., Lambert, P., 2006, MNRAS, 369, 1281

\bibitem[]{} Balmforth N.J. 1992a, MNRAS, 255, 603

\bibitem[]{} Balmforth N.J. 1992b, MNRAS, 255, 639

\bibitem[]{} Barnes, S. A., 2003, ApJ, 586, 464

\bibitem[]{} Basu S., Christensen-Dalsgaard, J., Monteiro, M.J.P.F.G.,
Thompson, M. J., 2001, in Proc. SOHO 10/GONG 2000 Workshop: Helio- and
Asteroseismology at the dawn of the millennium, ESA SP-464,
eds. A. Wilson, p.~407

\bibitem[]{} Ballot J., Turck-Chi\'eze S., Garc\'ia R. A., 2004, A\&A,
  423, 1061

\bibitem[]{} Basu S., Mazumdar A., Antia H. M., Demarque P., 2004,
  MNRAS, 350, 277

\bibitem[]{} Bedding, T.~R., Kjeldsen, H., Arentoft, T., et al., 2007,
ApJ, 663, 1315

\bibitem[]{} Bedding, T. R., 2011, in: `Asteroseismology: Canary
  Islands Winter School of Astrophysics 2011', 2011, vol 22,
  ed. P.~L.~Pall\'e, Cambridge University Press

\bibitem[]{} B\"ohm-Vitense, E., 2007, ApJ, 657, 486

\bibitem[]{} Campante, T. L., Handberg, R., Mathur, S., et al., 2011,
A\&A, in the press

\bibitem[]{} Chaplin W. J., Elsworth Y., Houdek G., New R., 2007,
MNRAS, 377, 17

\bibitem[]{} Chaplin, W. J., Houdek, G., Appourchaux, T., Elsworth,
  Y., New, R., Toutain, T., 2008, A\&A, 483, 43

\bibitem[]{} Chaplin, W. J., Appourchaux, T., Elsworth, Y., et al.,
  2010, ApJ, 713, L169

\bibitem[]{} Chaplin, W. J., Kjeldsen, H., Christensen-Dalsgaard, J.,
  et al., 2011a,  Sci, 332, 205

\bibitem[]{} Chaplin, W. J., Bedding, T. R., Bonanno, A., et al., 2011b,
ApJ, 732, 5L

\bibitem[]{} Chaplin, W. J., 2011, in: `Asteroseismology: Canary
  Islands Winter School of Astrophysics 2011', 2011, vol 22,
  ed. P.~L.~Pall\'e, Cambridge University Press

\bibitem[]{} Christensen-Dalsgaard, J., Bedding, T. R., Kjeldsen, H.,
  1995, ApJ, 443, L29

\bibitem[]{} Cunha, M. S., Aerts, C., Christensen-Dalsgaard, J., 2007,
A\&ARv, 14, 217

\bibitem[]{} Cunha, M. S., Metcalfe, T. S., 2007, ApJ, 666, 413

\bibitem[]{} Cunha, M. S., Brand\~ao, I. M., 2011, A\&A, 529, 10

\bibitem[]{} Deheuvels, S., Michel, E., 2010, Ap\&SS, 328, 259

\bibitem[]{} Deheuvels, S., Bruntt, H., Michel, E., et al., 2010,
  A\&A, 515, 87

\bibitem[]{} Do\u{g}an, G., Brand\~ao, I. M., Bedding, T. R.,
  Christensen-Dalsgaard, J., Cunha, M. S., Kjeldsen, H., 2010, Ap\&SS,
  328, 101

\bibitem[]{} Garc\'ia, R. A., Mathur, S., Salabert, D., Ballot, J.,
  R\'egulo, C., Metcalfe, T. S., Baglin, A., 2010, Sci, 329, 1032

\bibitem[]{} Gilliland, R. L., Brown, T. M., Christensen-Dalsgaard, J.,
  et al., PASP, 2010, 122, 131

\bibitem[]{} Giradi, L., Bertelii, G., Bressnan, A., Chiosi, C.,
Gr\"onewegen, M. A. T., Marigo, P., Salasnich, B., Weiss, A., 2002,
A\&A, 391, 195

\bibitem[]{} Giradi, L., Grebel, E., K., Odenkirchen, M., Chiosi, C.,
2004, A\&A, 422, 205

\bibitem[]{} Gizon L., Solanki, S. K., 2003, ApJ, 589, 1009

\bibitem[]{} Goldreich, P., Keeley, D. A., 1977, ApJ, 212, 243

\bibitem[]{} Gough, D. O., 1987, Nature, 326, 257

\bibitem[]{} Grundahl, F., Christensen-Dalsgaard, J., Arentoft, T.,
  Frandsen, S., Kjeldsen, H., Joergensen, U. G., Kjaegaard, P., 2009,
  CoAst, 158, 345

\bibitem[]{} Hekker, S., Broomhall, A.-M., Chaplin, W. J., Elsworth,
Y., Fletcher, S. T., New, R., Arentoft, T., Quirion, P.-O., Kjeldsen,
H., 2010, MNRAS, in the press

\bibitem[]{} Houdek, G., Balmforth, N. J., Christensen-Dalsgaard, J.,
Gough, D. O., 1999, A\&A, 351, 582

\bibitem[]{} Houdek, G., Gough, D. O., 2007, MNRAS, 375, 861

\bibitem[]{} Huber, D., Stello, D., Bedding, T. R., et al., 2009,
CoAst, 160, 74

\bibitem[]{} Huber, D., Bedding, T. R., Stello, D., et al., 2011, ApJ,
in the press

\bibitem[]{} Karoff, C., Metcalfe, T. S., Chaplin, W. J., Elsworth,
Y., Kjeldsen, H., Arentoft, T., Buzasi, D., 2009, MNRAS, 399, 914

\bibitem[]{} Kjeldsen, H., Bedding, T. R., Viskum, M., Frandsen,
   S. 1995, AJ, 109, 1313

\bibitem[]{} Marigo, P., Girardi, L., Bressan, A. et al. 2008, A\&A,
482, 883

\bibitem[]{} Mathur, S., Handberg, R., Campante, T. L., et al., 2011,
ApJ, 733, 95

\bibitem[]{} Mathur, S., Garc\'\i a, R. A., R\'egulo C., et al., 2010,
A\&A, in the press

\bibitem[]{} Mazumdar, A., Basu, S., Collier, B. L., Demarque, P.,
2006, MNRAS, 372, 949

\bibitem[]{} Metcalfe T. S., Dziembowski W. A., Judge P. G., Snow M.,
2007, ApJ, 379, 16

\bibitem[]{} Metcalfe, T. S., Monteiro, M. J. P. F. G., Thompson,
M. J., et al., 2010a, ApJ, 723, 1583

\bibitem[]{} Metcalfe, T. S., Basu, S., Henry, T. J.,
  Soderblom, D. R., Judge, P. G., Kn\"olker, M., Mathur, S., Rempel,
  M., 2010b, ApJ, 723, 213

\bibitem[]{} Michel, E., Baglin, A., Auvergne, M., et al., 2008, Sci,
322, 558

\bibitem[]{} Monteiro, M. J. P. F. G., Christensen-Dalsgaard, J.,
Thompson, M. J., 2000, MNRAS, 365, 165

\bibitem[]{} Mosser, B., Appourchaux, T., 2009, A\&A, 508, 877

\bibitem[]{} Osaki, Y., 1975, PASJ, 27, 237

\bibitem[]{} Roxburgh, I. R., Vorontsov, S. V., 2003, ApSS, 284, 187

\bibitem[]{} Roxburgh I. W., 2009, A\&A, 493, 185

\bibitem[]{} Samadi, R., Goupil, M.-J., 2001, A\&A, 370, 136

\bibitem[]{} Silva-Aguirre, V., Chaplin, W. J., Ballot, J., et al.,
2011a, ApJ, in the press

\bibitem[]{} Silva-Aguirre, V., Ballot, J., Serenelli, A. M., Weiss,
  A., 2011b, A\&A, 529, 63

\bibitem[]{} Soriano M., Vauclair S., 2008, A\&A, 488, 975

\bibitem[]{} Verner G. A., Chaplin W. J., Elsworth Y., 2006, ApJ, 638,
440

\bibitem[]{} Verner, G. A., Elsworth, Y., Chaplin, W. J., 2011, MNRAS,
tmp.892

\bibitem[]{} White, T. R., Bedding, T. R., Stello, D., et al., 2011a,
ApJ, submitted

\bibitem[]{} White, T. R., Bedding, T. R., Stello, D., et al., 2011b,
ApJ, in the press

\end{thebibliography}
\end{document}